\documentclass[%
aip,
 bmf,
 amsmath,amssymb,
 reprint,%
]{revtex4-1}

\usepackage{graphicx}
\usepackage{dcolumn}
\usepackage{bm}

\usepackage[utf8]{inputenc}
\usepackage[T1]{fontenc}

\begin{document}

\preprint{AIP/123-QED}

\title[Response of endothelial cells to a reduction in shear stress stimulation]{The effect of shear stress reduction on endothelial cells: a microfluidic study of the actin cytoskeleton}

\author{Mehdi Inglebert}
\affiliation{ 
Univ. Grenoble Alpes, CNRS, LIPhy, 38000 Grenoble, France
}%
\author{Laura Locatelli}
\affiliation{ 
Dept Biomedical and Clinical Sciences L. Sacco, Univ di Milano, Milano I-20157 Italy
}%
\author{Daria Tsvirkun}%
\affiliation{ 
Univ. Grenoble Alpes, CNRS, LIPhy, 38000 Grenoble, France
}%
\affiliation{ 
Belozersky Institute of Physico-chemical Biology, Lomonosov Moscow State University, Moscow, 119991, Russia
}%
\affiliation{ 
Federal Research and Clinical Center of Physical-Chemical Medicine of Federal Medical Biological Agency, Moscow, 119435, Russia
}%
\author{Priti Sinha}
\affiliation{ 
Univ. Grenoble Alpes, CNRS, LIPhy, 38000 Grenoble, France
}%
\author{Jeanette A. Maier}
\affiliation{ 
Dept Biomedical and Clinical Sciences L. Sacco, Univ di Milano, Milano I-20157 Italy
}%
\author{Chaouqi Misbah}
\author{Lionel Bureau}
\email{lionel.bureau@univ-grenoble-alpes.fr}
\affiliation{ 
Univ. Grenoble Alpes, CNRS, LIPhy, 38000 Grenoble, France
}%

\date{\today}

\begin{abstract}
Reduced blood flow, as occurring in ischemia or resulting from exposure to microgravity such as encountered in space flights, induces a decrease in the level of shear stress sensed by the endothelial cells forming the inner part of blood vessels. In the present study, we use a microvasculature-on-a-chip device in order to investigate {\it in vitro} the effect of such a reduction in shear stress on shear-adapted endothelial cells. We find that, within one hour of exposition to reduced wall shear stress, human umbilical vein endothelial cells undergo a reorganization of their actin skeleton, with a decrease in the number of stress fibers and actin being recruited into the cells' peripheral band, indicating a fairly fast change in cells' phenotype due to altered flow.
\end{abstract}

\maketitle

\section{\label{sec:intro}Introduction}

The vascular endothelium is the monolayer of specialized cells that lines the inner surface of blood vessels. It is essential to maintaining blood and vascular homeostasis through its functions in biochemical signaling, in vasomotion, and as a selective barrier controlling transport across and interactions with the vessel walls \cite{Chiu:2011iw,Davies:2008ia}. Being in direct contact with blood flow, Endothelial Cells (EC) have been shown to adapt to the hemodynamic forces to which they are exposed \cite{Chiu:2011iw,Chien:2007dp,Davies:2008ia}. 

Over the past forty years, a breadth of studies have investigated the influence of fluid shear stress on the behavior and function of endothelial cells, and demonstrated the importance of the level and pattern of shear stress on {\it e.g.} EC morphology  \cite{Resnick:2003dc,Malek:1996}, proliferation \cite{Levesque:1990ih,Lin:2000kp}, gene expression \cite{Malek:1995bj}, or the development of vascular pathologies \cite{Chiu:2011iw}.

One of the first observations of EC adaptation to shear forces produced by blood stream is their elongation and preferred orientation along the flow direction \cite{Levesque:1985ia}. The establishment of this so-called atheroprotective phenotype  \cite{Traub:1998dn} involves multiple players. The glycocalyx, or endothelial surface layer (ESL), is the primary structure exposed to blood flow \cite{Reitsma:2007ei,Florian:2003eb,Thi:2004df,Zeng:2014ji}. This polymer-rich  hydrated layer attached to the apical surface of EC acts, along with others \cite{Davies:1995bc,Tzima:2005be,Rizzo:2003by}, as a sensor that transmits, through transmembrane proteins, the mechanical stimulus to the actin cytoskeleton of the cells \cite{Tarbell:2006er,Thi:2004df}. This triggers a reorganization of the actin network, associated with a remodeling of intercellular junctions and focal adhesions at the basal side of EC \cite{Noria:1999ge,Thi:2004df,Davies:2008ia,Noria:2004cz}. A salient feature of the cytoskeleton of shear-adapted EC is the presence of so-called stress fibers composed of body-spanning elongated bundles of filamentous actin, aligned in the flow direction and contributing to the overall shape and stability of the atheroprotective phenotype \cite{Franke:1984gc,Galbraith:1998ci,Noria:2004cz}. 

While pioneer {\it in-} or {\it ex vivo} studies \cite{Wong:1983jk,Nehls:1991im,Langille:1991ee,Kim:1989gi,Walpola:1993wi} have set the basis of the above picture regarding the shear-dependent organization of the actin network, most of the current knowledge in the field comes from {\it in vitro} experiments performed using {\it e.g.} parallel plate flow chambers in which 2-dimensional ({\it i.e.} not laterally confined) confluent cultured cells are submitted to well-controlled hydrodynamic forces \cite{Chiu:2011iw}. Recently, microfluidic tools have been combined with EC culture in order to investigate the cellular response to flow in microvessel-mimicking devices \cite{vanderMeer:2010bo,Zheng:2017kn,TovarLopez:2019gy}. Such  studies have been performed over a range of endothelial cell types (from Human Umbilical Vein \cite{Blackman:2002ek}, Human Pulmonary Artery \cite{Ting:2012iv}, Bovine Pulmonary artery \cite{Birukov:2002ev}, Porcine Aorta \cite{Noria:2004cz}, Bovine Aorta \cite{Girard:1995hn,Malek:1996}, Rat fat pad capillaries \cite{Thi:2004df}), thus covering the response of cells coming from various part of the vascular tree. {\it In vitro} investigations have thus put forward the impact on actin organization of parameters such as the level of steady shear stress applied to EC \cite{Levesque:1985ia,Malek:1999jk}, the magnitude of stress gradients \cite{Sakamoto:2010gl}, the duration of the applied shear stimulus \cite{Girard:1995hn,Noria:2004cz,Noria:1999ge,Birukov:2002ev,Galbraith:1998ci}, the pulsatility of the flow \cite{Blackman:2002ek,Mohammed:2019gy}, or flow perturbations such as recirculation or stagnation \cite{Chiu:1998ib,Ting:2012iv,TovarLopez:2019gy}. 

However, it is interesting to note that in all these {\it in vitro} studies, EC are either (i) cultured under ``no flow'' conditions until confluence, and then placed under hydrodynamic shear, or (ii) cultured under flow from the moment they are plated into the flow device, but experiencing the same level (high or low) or pattern (laminar, recirculating, graded or pulsating) of shear stress {\it from the start} of the culture. Most studies therefore investigated the response of EC when flow is ``switched on'', or of EC submitted to specific shear conditions applied steadily from ``time zero''. Strikingly enough, only a very limited number of works have addressed the question of the response of already shear-adapted EC to a loss of shear stress stimulation \cite{Fisher:2001df,Manevich:2001hf,Leemreis:2006fh,Milovanova:2008dh,Walpola:1993wi,Kim:1989gi}, as would occur for instance in occlusive microvascular events, or as a result of altered hemodynamics such as encountered in space flights \cite{Zhang:2001fu,Hughson:2018ic} or in ground simulations of microgravity conditions \cite{Delp:2000jb}. Therefore, the question of how endothelial cells that experienced a physiological shear stress for a long period of time react to a reduction of the applied shear remains largely open.

This issue fostered the present work, in  which we study the impact of a 5-fold decrease of shear stress on the actin network organization of EC that were previously shear-adapted. We do so by using a microvasculature-on-a-chip where endothelial cells are grown to confluence under flow in a network of microchannels \cite{Tsvirkun:2017ea}. We observe that, when cultured during 96h under physiologically relevant shear stress (0.4 Pa), endothelial cells elongate in the flow direction and form long and thick stress fibers. Starting from such a well-known atheroprotective phenotype, a decrease in shear stress results, within one hour, in a  disorganization of the stress fibers, combined with an increase in peripheral actin and a thickening of cellular body. We also observe that a specific degradation of the glycocalyx, targeted to sialic acid residues, results in no change in the amount and orientation of stress fibers of shear-adapted EC.

\section{\label{sec:MatMet}Materials and methods}

\subsection{\label{microflu}Microfluidics} 

\begin{figure}[htpb]
\includegraphics[width=\columnwidth]{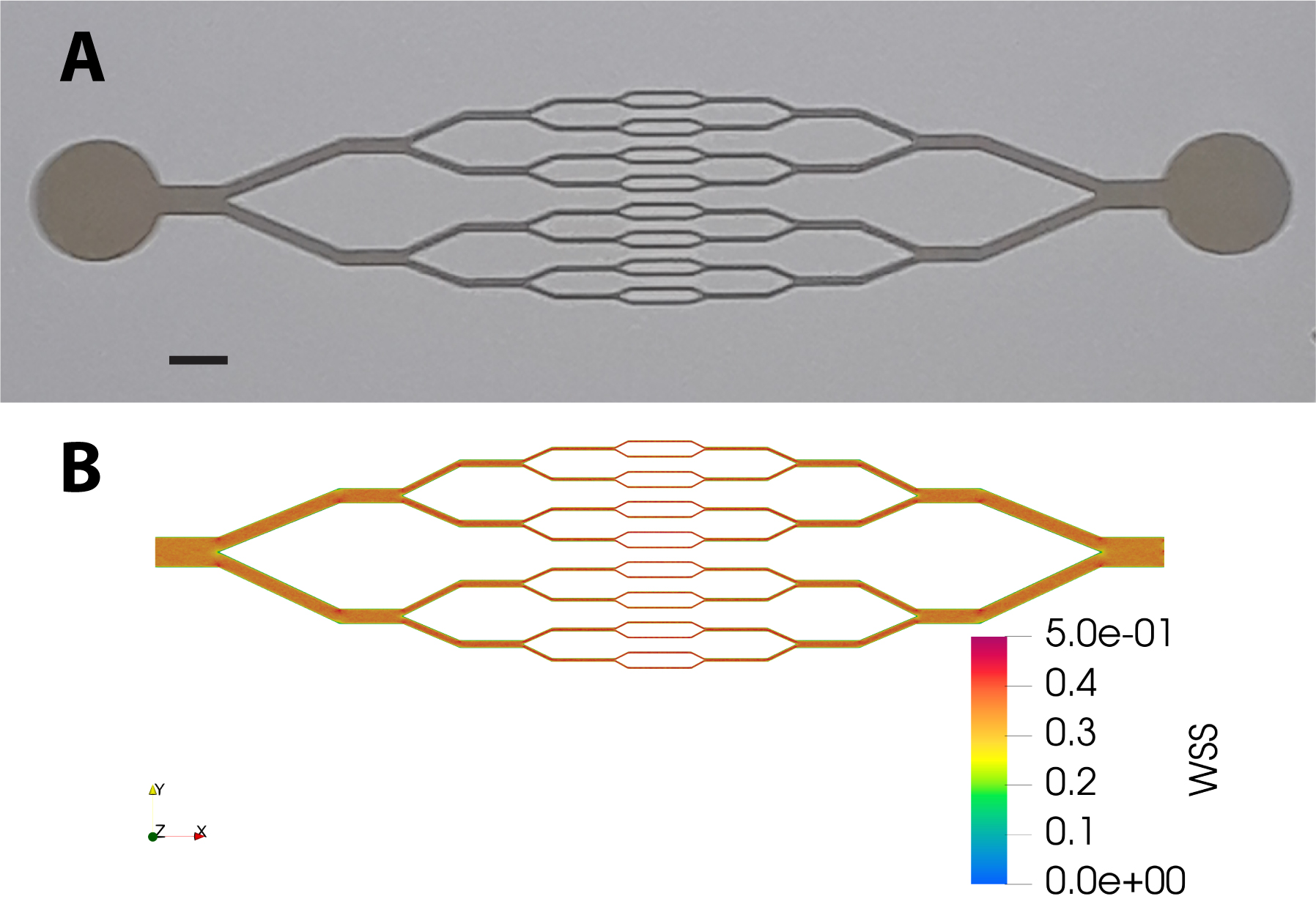}
\caption{\label{fig:fig1} (A) Picture of the resin mold showing the geometry of the microfluidic network (scale bar : 1 mm). (B) Map of the wall shear stress (WSS, in Pa) calculated by Computer Fluid Dynamics for an imposed flow rate of $1\,\mu$L.min$^{-1}$.}
\end{figure}

Microchannel networks displaying diverging/converging geometry (see Fig. \ref{fig:fig1}A) were fabricated using a standard soft lithography technique. A master mold of the network was obtained from a positive photoresist (SU8, Gerseltec) cast onto a silicon wafer and exposed to UV light through a quartz-chromium photomask. PDMS (Sylgard 184, Dow Corning) was cast onto the mold and cured for 2 hours at 65$^{\circ}$C. A glass coverslip ($\#$0, thickness $\approx$100 $\mu$m) was used as the bottom part of the microchip and was permanently sealed to the PDMS upper part after exposure of the surfaces of both elements to oxygen plasma (PDC-32G-2, Harrick). The thickness of the photoresist on the master mold was set such that the series of 16 parallel microchannels in the central part of the network had a square cross-section of 30$\times$30 $\mu$m$^2$. The circuit was connected with silicone tubing to a fluid reservoir at the inlet, and at the outlet to a 1 mL syringe installed on a high precision syringe pump (Legato 110, KD Scientific) used in withdraw mode at imposed flow rate.

The design of our circuit, in which the height of the channels is uniform ($30\,\mu$m) while their width is divided (resp. multiplied) by 2 at each branching (resp. converging)  point, is such that for a given flow rate the fluid shear stress exerted at the bottom and top walls is expected to be homogeneous, irrespective of the channel size. In order to confirm this, we have performed tridimensional numerical modeling of a laminar flow within our network geometry, using  Computer Fluid Dynamics (Simscale GmbH). As shown in Fig. \ref{fig:fig1}B, numerical simulations indicate that wall shear stress is indeed fairly uniform across the circuit: at an imposed flow rate of $1\,\mu$L.min$^{-1}$, we compute a wall shear stress of $0.4\pm0.05$ Pa at the bottom or top surfaces.

\subsection{\label{cell}Cell culture and staining}  

\begin{figure*}[ht!]
\includegraphics{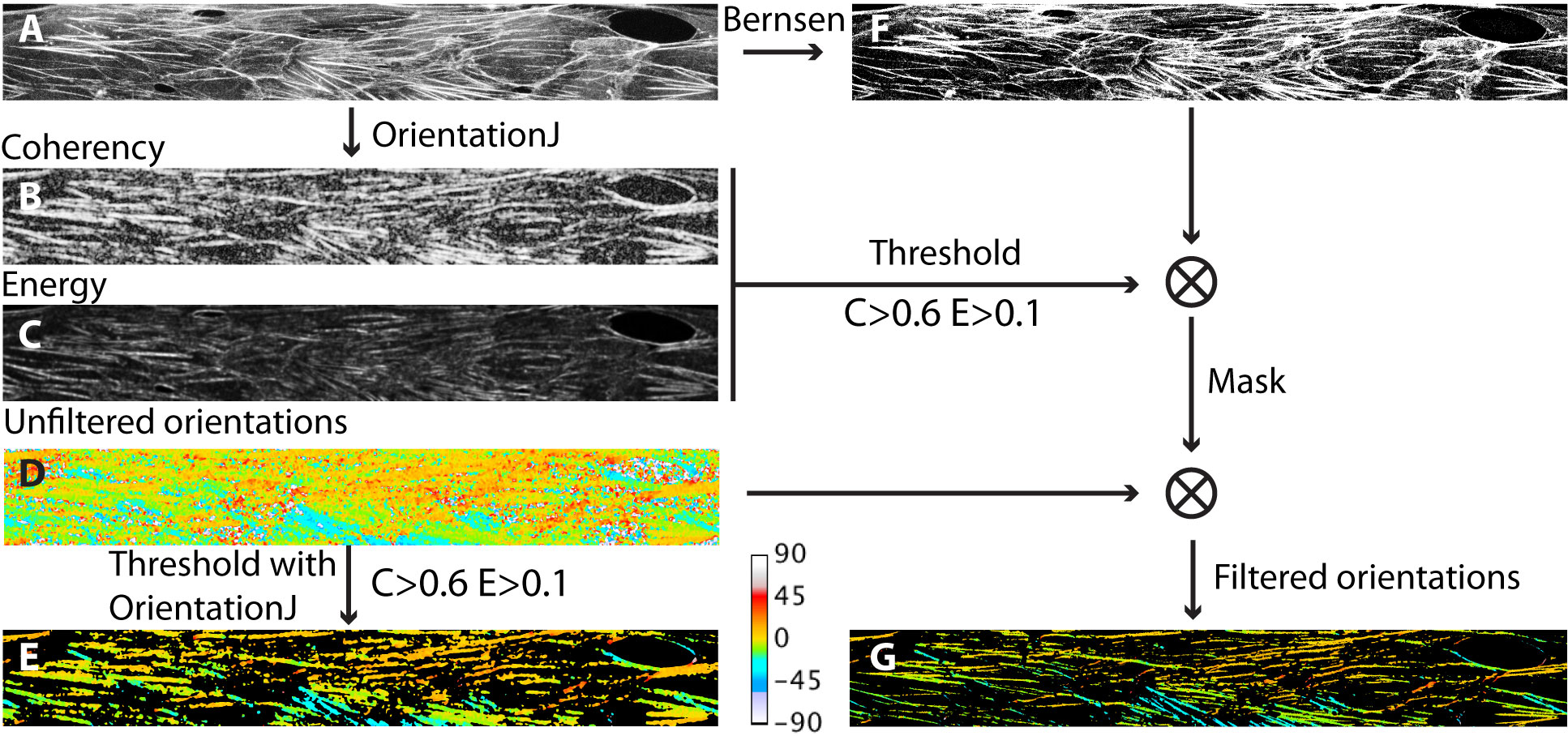}
\caption{\label{fig:fig2} Flowchart of image analysis: starting from a F-actin image (A), the OrientationJ plugin outputs Coherency (B), Energy (C), and Orientation (D) maps. OrientationJ allows filtering the orientation map in order to retain only pixels involved in regions having coherency and energy values larger than a given threshold ($E>0.1$ and $C>0.6$ here) (E). As an alternative to the latter step, we apply a local thresholding scheme (Bernsen) to the initial image (F) and multiply this segmented image by the thresholded maps (B) and (C) in order to create a binary mask. We then use such a mask in order to discard irrelevant pixels from the raw orientation map (D), and obtain the orientation map (G), with better spatial resolution than (E). Colors on images (D), (E) and (G) code for the local angle, according to the lookup table displayed at the bottom of the figure.}
\end{figure*}

Single donor Human Umbilical Vein Endothelial Cells (HUVEC) were obtained from Promocell GmbH (Germany) and cultured in endothelial cells growth medium supplemented by growth factors (Supplement Mix, Promocell GmbH, Germany).

Before cell seeding, microchannels were activated by exposure to oxygen plasma and coated with human fibronectin (50 $\mu$g/mL, Corning) for 30 min at 37 $^{\circ}$C. Endothelial cells were suspended in culture medium at a concentration of 10$^7$ cells/mL and filtered through a nylon cell strainer with pore diameter of 40 $\mu$m. The cell suspension was injected into the microchannel network with a pressure controller (OB-1 MK2, Elveflow, France). Consecutive steps of injection of a few seconds and sedimentation periods of several minutes were performed until about 50 \% of the network surface was covered by cells. 

Cells were then cultured at 37 $^{\circ}$C and 5\% CO$_{2}$ under a constant flow of supplemented medium for one night at 0.6 $\mu$L/min and then maintained at 1$\mu$L/min for four days in order for cells to form a confluent layer and adapt to shear. As mentioned in the previous section, a flow rate of 1$\mu$L/min corresponds to a wall shear stress of $\sim$0.4 Pa, representative of physiological conditions in {\it e.g.} post-capillary venules \cite{Sheikh:2003bk}. 

After four days of culture under 0.4 Pa, the shear-adapted cells were exposed to a reduced flow rate of 0.2 $\mu$L/min, corresponding to a reduced wall shear stress of 0.08 Pa. Reduced flow was applied for a duration of 1 or 6 hours, after which cells were fixed by perfusing the channels with a solution of 4\% paraformaldehyde in PBS for 30 min at 37 $^{\circ}$C, and stained for observation.

As controls, HUVEC were also cultured under (i) static conditions into fibronectin-coated glass-bottom Petri dishes until reaching confluence, and (ii) reduced flow conditions (wall shear stress 0.08 Pa) applied from the start of the experiment, during four days of culture after seeding in the channels.

All experiments were performed in duplicates.

For confocal microscopy we used the following dyes: (i) Hoechst 33342 (Molecular Probes) for nuclei staining, (ii) Phalloidin TRITC (Sigma-Aldrich) for F-Actin staining and (iii) Wheat Germ Agglutinin (WGA) Alexa Fluor 488 Conjugate (5 $\mu$g/ml, Molecular Probes) for staining the glycocalyx. The dyes were used in accordance with manufacturer's specification.  

For glycocalyx modification, cells were exposed to Neuraminidase from Clostridium perfringens (1U/ml, Sigma-Aldrich) in a CO$_2$ incubator. Neuraminidase was used at 150 mUI/mL and perfused at $1\,\mu L$/min for 3 hours in non-supplemented endothelial cells culture medium. Glycocalyx was stained by WGA-Alexa 488 after enzymatic modification and fluorescence intensity was estimated with and without Neuraminidase action.

\subsection{\label{microscopy}Image acquisition and processing}  

Confocal fluorescence microscopy was performed on an inverted microscope. XYZ image stacks were obtained in raster mode using a Zeiss LSM710 module and a 40$\times$/NA1.3 oil-immersion objective, with a lateral size of 1024$\times$200 pixels, a lateral resolution of 0.225 $\mu$m/pixel, and a Z-slice spacing of 0.46 $\mu$m.

The acquired image stacks were processed and analyzed using the ImageJ open-source platform and its built-in plugins.
3-dimensional XYZ image stacks were treated as follows. A point spread function (PSF) of the microscope was generated numerically with ImageJ, using the ``Diffraction PSF 3D'' plugin fed with the experimental characteristics of our imaging system (i.e. refraction index of immersion liquid, objective numerical aperture, lateral magnification and z slice spacing of the stack). We have further checked for the consistency of such a generated PSF by using it to deconvolve images of fluorescent latex beads of known diameter (500 nm) and ensuring
that the bead size measured from the deconvolved images agreed with the expected one. PSF deconvolution was then performed on image stacks of the stained cells using the ``Iterative deconvolve 3D'' plugin. Z-projections of deconvolved image stacks were then made, summing the intensity of each of the slices spanning the EC monolayer adhered on the bottom surface of the channels.

\subsection{\label{orient}Quantification of filamentous actin structures}  

From Z-projected images obtained as described above, F-Actin structures were characterized with the ImageJ plugin ``OrientationJ'' \cite{Rezakhaniha:2011jd}, using the ``Gaussian gradient'' built-in method with a window size of 2 pixels. 

The OrientationJ plugin computes the structure tensor from the intensity gradient of an image, and attributes to each pixel in the image a pair of values, between 0 and 1, corresponding to the local values of the (normalized) energy, E, and coherency, C, of the structure tensor. Low values of E correspond to regions in the image where intensity is uniform, while large values indicate regions having one or multiple orientations. C is a measure of the local degree of anisotropy, with values closer to 1 corresponding to regions of stronger anisotropy. 

For each experimental conditions, we compute E and C maps of the F-actin images obtained (such as illustrated in Fig. \ref{fig:fig2}A-C), and build a final map where each pixel contains the value of $\sqrt{EC}$. From such maps we determine the distribution of $\sqrt{EC}$ by building histograms over 50 bins of $\sqrt{EC}$ spanning the range $0-1$, where we compute, for each bin, a frequency corresponding to the number of pixels displaying a given value of $\sqrt{EC}$ divided by the total number of pixels in the image. Plotting the distribution of  $\sqrt{EC}$ allows us assessing, for a given cell culture condition, the fraction of image pixels  involved in non-uniform and strongly elongated F-actin structures.

From the computation of the structure tensor, OrientationJ also provides a map of the local orientation. The plugin associates an orientation to each pixel in the image, including in regions of low E and C where this quantity is not meaningful. Such an orientation map can then be filtered by thresholding on E and C values in order to retain only the orientation values of pixels that are actually part of anisotropic structures. Doing so using the built-in thresholding feature of OrientationJ, we observe that, with our type of images, the filtered orientation maps display a spatial resolution that is clearly lower than the initial image, with blurred features and inclusion of pixels that are actually outside of the fibers (see Fig. \ref{fig:fig2}). We could not find a set of OrientationJ parameters that would allow us to increase further the resolution of the orientation map without losing part of the elongated structures that are visible in the original image. As an alternative to the built-in thresholding feature of the plugin, we find that the following procedure performs better in terms of final resolution: starting from the input image, we run orientationJ to obtain E, C and unfiltered orientation maps. We then threshold the E and C images (at the same levels used in the built-in function of the plugin, namely $C>0.6$ and $E>0.1$ in the present work) and binarize them. In parallel, we binarize the input image using the local threshold Bernsen method implemented in ImageJ. We then multiply the binarized E, C and Bernsen image in order to create a binary mask that we use to filter the orientation map. A comparison of the built-in and custom-made thresholding is provided in Fig. \ref{fig:fig2}. 

The orientation maps thus produced are then used to compute angular distributions. In the following, angular distributions are computed over 30 bins of width 3$^{\circ}$, spanning the range between 0$^{\circ}$ and +90$^{\circ}$, with 0$^{\circ}$ being the flow direction, and where we count in the same bin the pixels displaying an angle of $\theta$ or $-\theta$. Frequencies in each bin correspond to the ratio of the number of pixels within an angular bin to the total number of pixels involved in oriented F-actin structures.

Distributions of $\sqrt{EC}$ and angles were computed from images taken in the central part of the networks, {\it i.e.} in the channels having a cross-section of $30\times30\,\mu$m$^2$. 

\subsection{\label{stat}Statistical analysis}

The number of analyzed images was $N=53$ (389 cells) for culture under flow at 0.4 Pa, $N=32$ (311 cells) for flow reduced at 0.08 Pa for 1h, $N=26$ (235 cells) for flow reduced at 0.08 Pa for 6h, $N=16$ (151 cells) for glycocalyx degradation, $N=14$ (141 cells) for culture under 0.08 Pa for 4 days, and $N=6$ (132 cells) for static condition.

For every experimental condition, $\sqrt{EC}$ and angular distributions were computed on each of the $N$ images taken. In what follows, we present such distributions as mean frequency $\pm\, 2$ s.e.m. (standard error of the mean), computed bin-by-bin over the $N$ images analyzed. When comparing the results for various experimental conditions, we take as statistically significant any difference between distributions larger than 4 s.e.m. ({\it i.e.} when the 95\% confidence intervals of the distributions do not overlap).

Other quantifications (fluorescence intensity or length measurements on images) were analyzed using t-tests at the 0.05 threshold, performed using Origin software. In what follows, we indicate by (*) datasets whose means are statistically different ($p<0.05$), and by (**) those that are not ($p>0.05$).

\section{\label{sec:Res}Results}

We label as follows the various conditions under which HUVEC were imaged:

Condition (i): after 4 days of culture under 0.4 Pa of wall shear stress,

Condition (ii): after 4 days of culture under 0.4 Pa of wall shear stress, followed by 1h under 0.08 Pa,

Condition (iii): after 4 days of culture under 0.4 Pa of wall shear stress, followed by 6h under 0.08 Pa,

Condition (v): after 4 days of culture under 0.08 Pa of wall shear stress,

Condition (iv): after 4 days of culture under 0.4 Pa of wall shear stress, followed by exposition to neuraminidase for 3h under 0.4 Pa,

Condition (vi): after 4 days of culture under static condition, in the absence of flow.

\subsection{\label{methodo} Comparison of conditions (i) and (vi) : 0.4 Pa {\it vs} static culture}  

\begin{figure}[htbp]
\includegraphics[width=1\columnwidth]{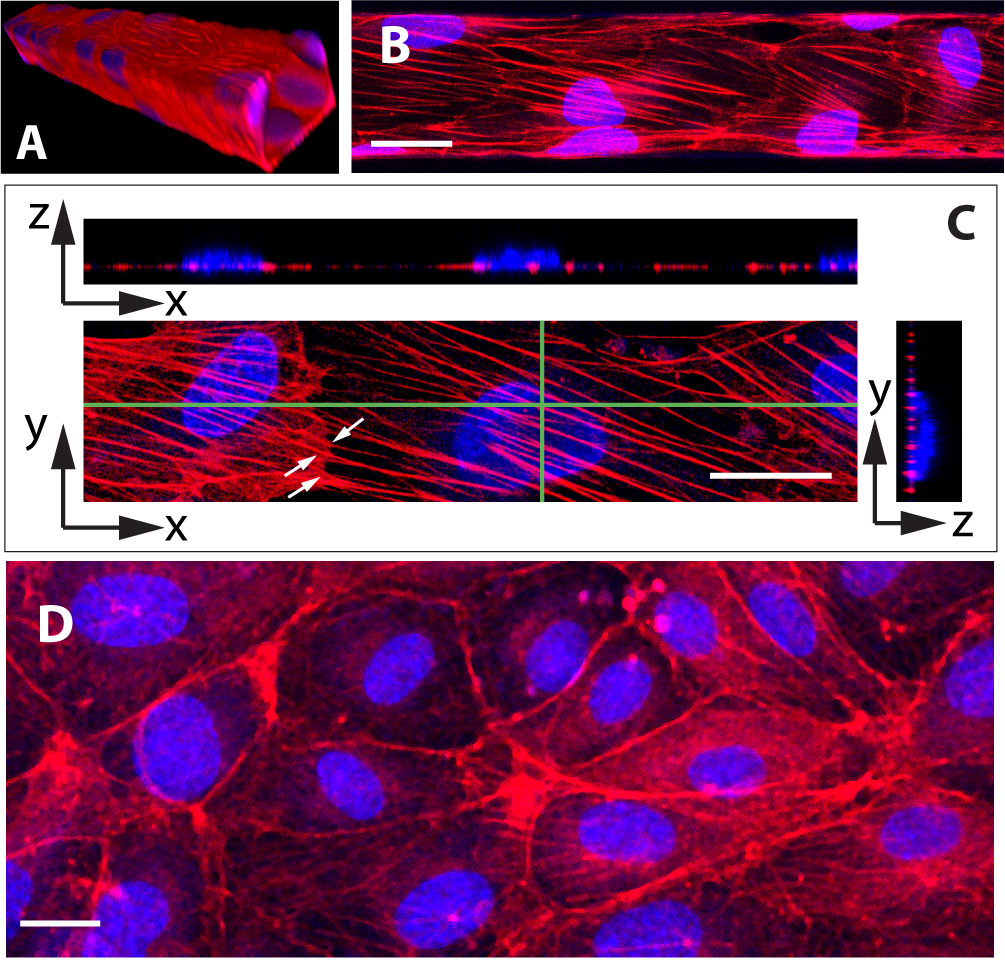}
\caption{\label{fig:fig3} (A) 3D rendering of images from a portion of channel covered by HUVEC, having a cross-section of $30 \times 30\, \mu$m$^2$ and a length of 212 $\mu$m (overlay of F-actin (red) and nuclei (blue)). (B) Image of the bottom surface of a microchannel ($30 \times 30\, \mu$m$^2$ in cross-section), showing the thick and oriented actin fibers displayed by HUVEC cultured under 0.4 Pa (flow direction from left to right. Scale bar 20 $\mu$m). (C) Image of adjacent cells (xy panel) showing actin fibers ``crossing'' the the cells' boundary (white arrows). Stress fibers are located at the basal side of the cells, as seen on the two orthogonal images showing fiber cross-sections running below the nuclei (xz and yz panels taken along the directions indicated by green lines on the xy view. Scale bar 20 $\mu$m) (D) Image of HUVEC cultured under static conditions, showing F-actin being mainly localized at the periphery of the cells, with only few and randomly oriented fibers (scale bar 20 $\mu$m).}
\end{figure}

As illustrated on Fig. \ref{fig:fig3}A, HUVEC cultured under 0.4 Pa of fluid shear stress for four days yield a confluent monolayer of cells covering the four walls of the microchannels. Under such conditions, we observe that cells exhibit many oriented actin stress fibers that cross the cell from side to side underneath the nucleus and seem to be attached to cell-cell junction as they continue end to end into the neighboring cell (see Fig. \ref{fig:fig3}B and C). In contrast to this, HUVEC cultured under static conditions rather display a marked deep peripheral actin band, with no or few randomly oriented stress fibers (Fig. \ref{fig:fig3}D). 

\begin{figure}
\includegraphics[width=\columnwidth]{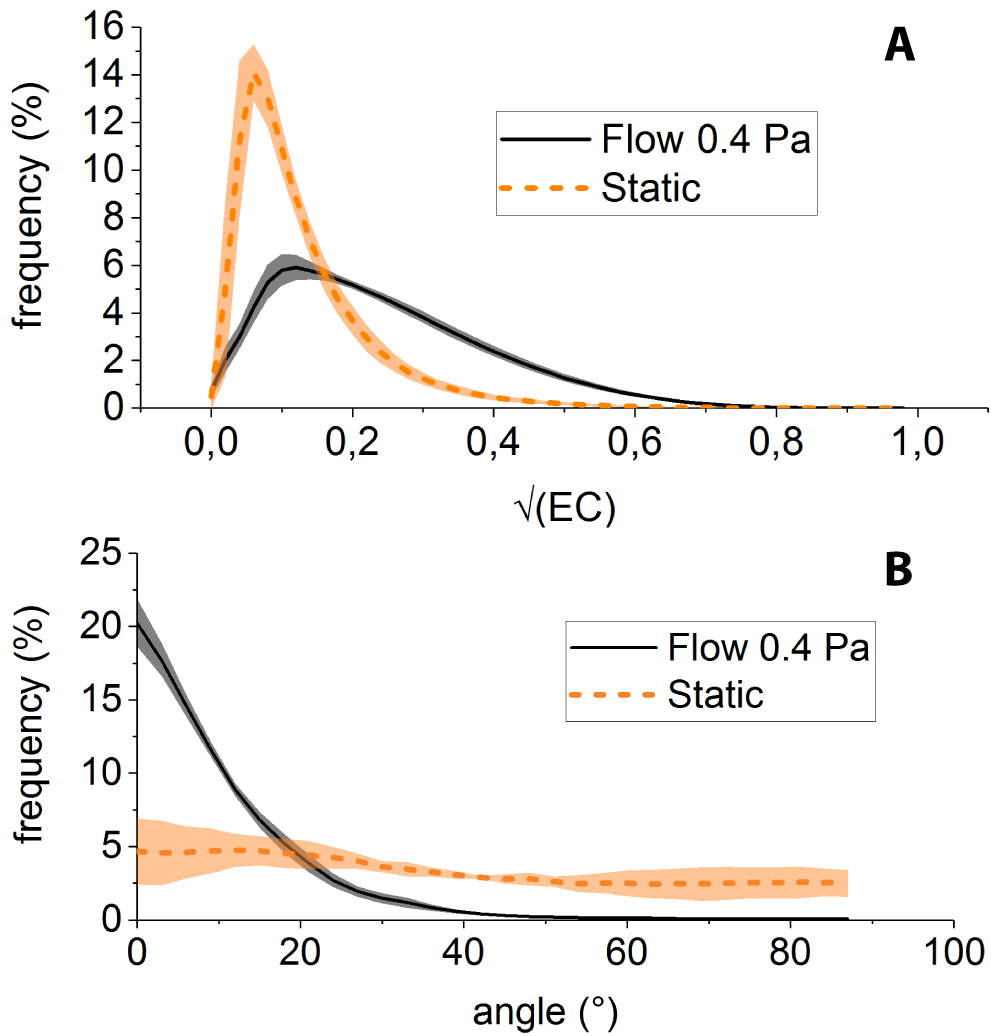}
\caption{\label{fig:fig4} (A) Distribution of $\sqrt{EC}$ for HUVEC cultured under 0.4 Pa in microfluidic channels (Black) and under static conditions (Orange). (B) Angular distribution of actin filamentous structures observed under flow (Black) and static conditions (Orange).}
\end{figure}

Applying the above-described procedure to F-actin images obtained under static and 0.4 Pa shear flow conditions, we observe that: (i) the distribution of $\sqrt{EC}$ is more populated towards large values ($\sqrt{EC} \geq 0.2$) for shear-adapted cells than for static condition (Fig. \ref{fig:fig4}A), consistent with more anisotropic structures ({\it i.e.} stress fibers) being present in cells cultured under fllow, and (ii) the angular distribution of such structures is sharply peaked about $0^{\circ}$ ({\it i.e.} along the flow direction) for shear-adapted cells (Fig. \ref{fig:fig4}B), indicating streamwise alignment of stress fibers, whereas the angular distribution of actin in static cultures is much flatter.

\subsection{\label{lowshear}Effect of reduced hydrodynamic stress after shear-adaption}

\begin{figure*}
\includegraphics[width=2\columnwidth]{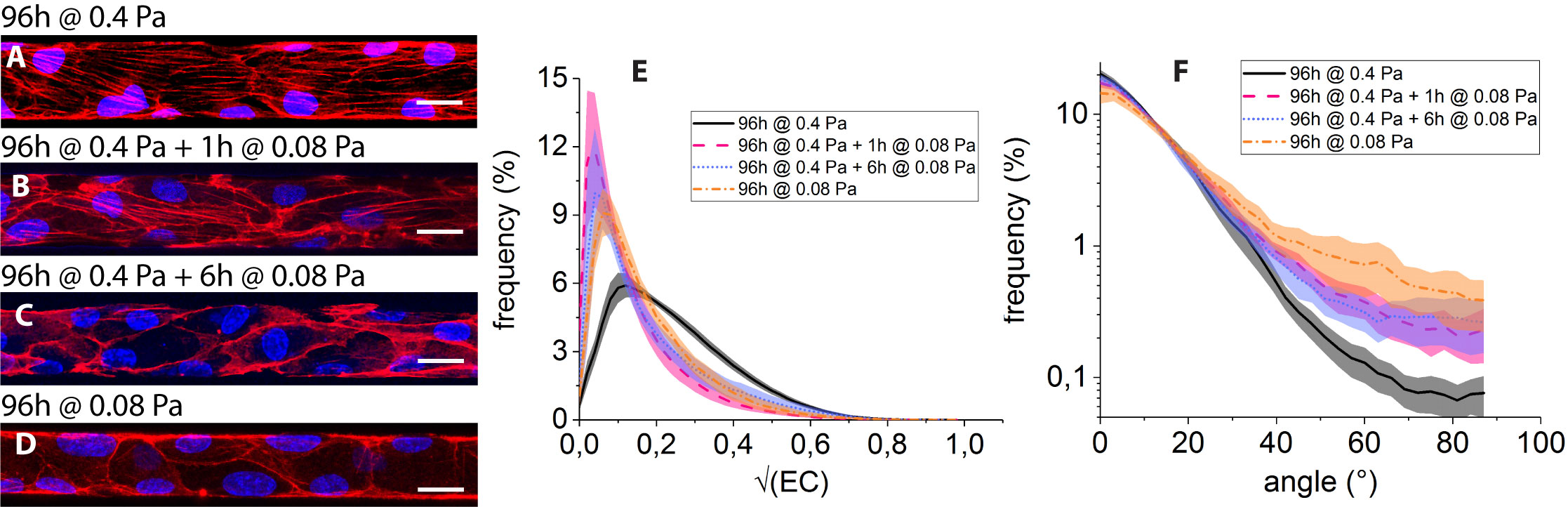}
\caption{\label{fig:fig5} Overlay images of cells' nuclei (blue) and F-actin (red) under (A) 96h at 0.4 Pa, (B), 96h at 0.4 Pa + 1h at 0.08 Pa, (C) 96h at 0.4 Pa + 6h at 0.08 Pa, and (D) 96h at 0.08 Pa. Panel (E): Distribution of $\sqrt{EC}$ for 96h at 0.4 Pa (black, continuous), 96h at 0.4 Pa + 1h at 0.08 Pa (magenta, dashed), 96h at 0.4 Pa + 6h at 0.08 Pa (parma, dotted), and 96h at 0.08 Pa (orange, dash-dotted). Shaded area around lines represent $\pm 95\%$ confidence interval for each dataset. (F) Angular distributions of actin filamentous structures, color code as in (E).}
\end{figure*}

The impact on the F-actin network of culture conditions (i) to (iv) is summarized on Fig. \ref{fig:fig5}. 

As already mentioned, culture under 0.4 Pa for 96h (condition (i)) results in the development of flow-aligned stress fibers (Fig. \ref{fig:fig5}A). In contrast to this, submitting the HUVEC, previously shear-adapted to 0.4 Pa for 96h, to 1h of shear stress lowered to 0.08 Pa (condition (ii)) leads to a marked decrease in the number of fibers and to a localization of F-actin at the periphery of the cells (Fig. \ref{fig:fig5}B). Such a trend is maintained after 6h of reduced flow (condition (iii), Fig. \ref{fig:fig5}C). 

In addition, we observe that HUVEC constantly submitted to 0.08 Pa for 96h of culture (condition (iv), Fig. \ref{fig:fig5}D) form a confluent monolayer in the microfluidic networks, and display an organization of F-actin presenting features that are qualitatively similar to those observed for conditions (ii) and (iii), with little or no stress fibers and a pronounced actin peripheral band.

The decrease in the number of filamentous structures under low shear stress can be seen on the distributions of $\sqrt{EC}$ (Fig. \ref{fig:fig5}E):  as observed under purely static condition, distributions computed for conditions (ii) to (iv) are not significantly different and all exhibit a peak at low values ($\sqrt{EC}\simeq0.1$), whereas that computed for condition (i) is significantly more populated $\sqrt{EC}>0.2$. 

Such a loss of actin fibers under 0.08 Pa of wall shear stress is accompanied by a broadening of the angular distributions, which exhibit tails beyond $40^{\circ}$ that are more pronounced than at 0.4 Pa (Fig. \ref{fig:fig5}F). While our image analysis procedure does not allow us to discriminate between the various contributions, the observed angular broadening under reduced flow can stem from both a disorientation of stress fibers and from the buildup of peripheral actin, both contributing to large angles in the distributions. 

Moreover, we see in Fig. \ref{fig:fig5}F that the angular distribution computed for condition (iv) lies above those computed for conditions (ii) and (iii). This may stem from the fact that HUVEC submitted to 1 or 6h of low shear stress after 96h at 0.4 Pa do however keep a flow-elongated shape, as seen in Fig. \ref{fig:fig5}B and C from the peripheral actin bands, while HUVEC submitted to permanent low shear do not exhibit flow alignment (Fig. \ref{fig:fig5}D), resulting in a broader angle distribution for condition (iv).

\begin{figure}
\includegraphics[width=1\columnwidth]{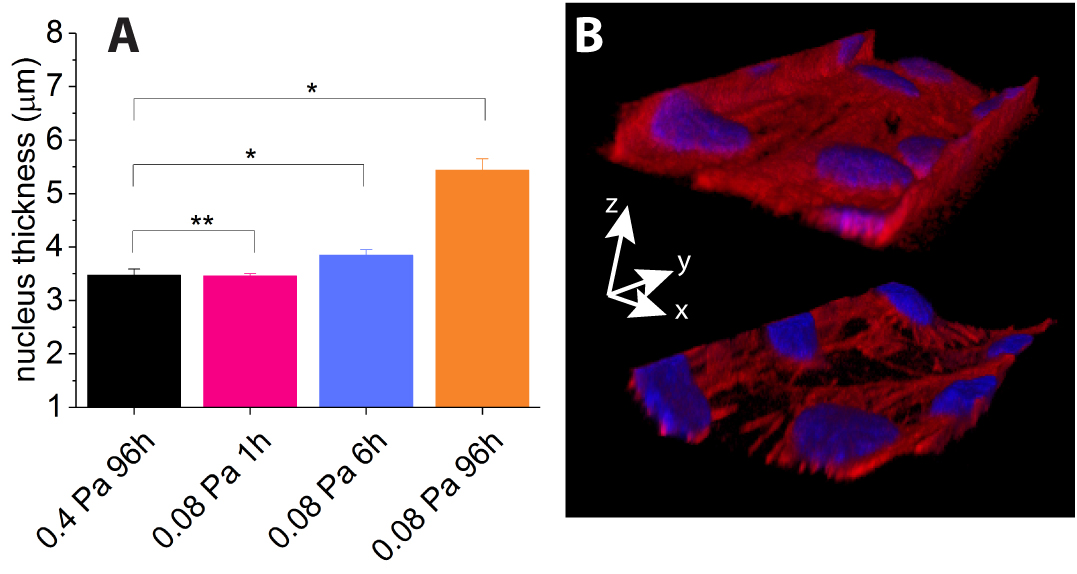}
\caption{\label{fig:fig6} (A) Mean nucleus thickness of HUVEC under indicated culture conditions: datasets for conditions (i) and (ii) are not statistically different (**, p>0.05), while datasets for conditions (iii) and (iv) exhibit a statistically significant increase compared to condition (i) (*, p<0.05.)(B) 3D reconstructions of the bottom part of a channel showing cells lying in the corners under 0.4 Pa (top image), while they tend to detach from corners and round up the effective lumen under 0.08 Pa (bottom image). The lateral channel size is 30 $\mu$m.}
\end{figure}

From our 3D confocal image sets, we have measured the maximum thickness of the nuclei of the cells, and notice an increase in thickness under prolonged reduced flow conditions, as shown on Fig. \ref{fig:fig6}A: while nucleus thickness between conditions (i) and (ii) are not statistically different, we measure a significant increase of thickness for conditions (iii) and (iv). 

In addition to this, we notice qualitatively that upon long exposure to low shear stress (>6h), 0.4Pa-shear-adapted HUVEC tend to detach from the microchannel walls, mainly from the corners, as illustrated on Fig. \ref{fig:fig6}B, while maintaining cell-cell contacts. 

\subsection{\label{lowshear}Effect of neuraminidase treatment}  

\begin{figure}
\includegraphics[width=1\columnwidth]{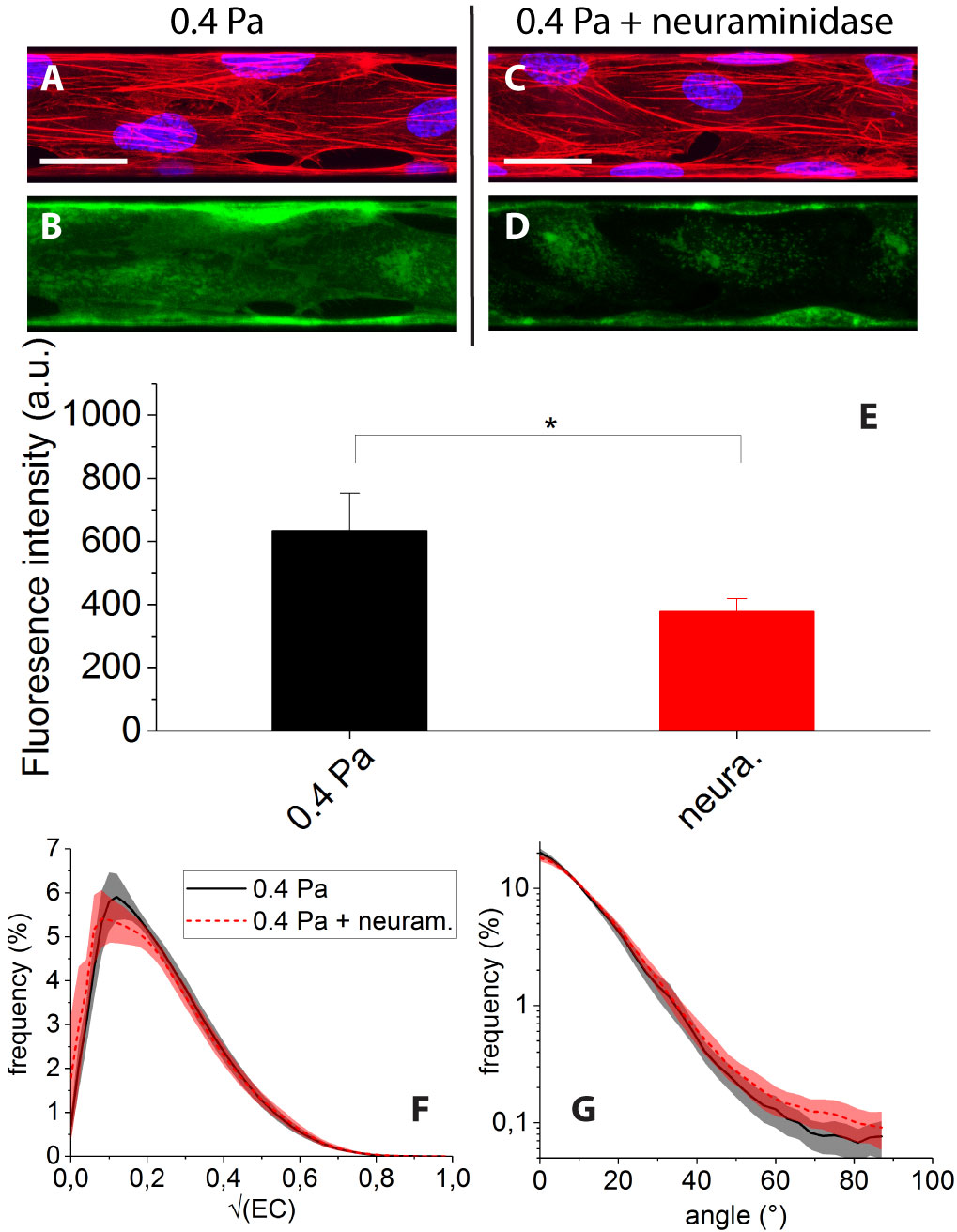}
\caption{\label{fig:fig7}  Overlay of nuclei and F-actin images of cells submitted to 0.4 Pa of shear stress,  before (A) and after (C) neuraminidase treatment (scale bar $20\,\mu$m). Corresponding WGA-Alexa fluorescence images taken before (B) and after (D) treatment show a marked decrease in intensity upon treatment, as confirmed by fluorescence intensity histograms (E). Distributions of $\sqrt{EC}$ (F) and fiber angles (G) do not significantly differ between control condition  (black continuous line) and degraded condition (red dashed line).}
\end{figure}

The effect of treating the cells with neuraminidase while submitting them to 0.4 Pa of shear stress is illustrated in Fig.  \ref{fig:fig7}. The enzymatic degradation of the glycocalyx results in a clear decrease of the WGA-Alexa fluorescence signal (see Fig. \ref{fig:fig7}C and D) which indicates actual removal of part of the andothelial surface layer. This can be quantified by a statistically significant 40$\%$ decrease of image mean intensity (Fig. \ref{fig:fig7}E). 

However we find that neuraminidase treatment does not induce any clear change of the F-actin organization: after enzymatic degradation (condition (v)), we observe cells displaying numerous and flow oriented stress fibers (Fig. \ref{fig:fig7}C), similarly to condition (i) (Fig. \ref{fig:fig7}A). Accordingly, distributions of $\sqrt{EC}$ and angles for conditions (i) and (v) are found to display no significant differences (Fig. \ref{fig:fig7}F and G).

\section{\label{sec:Discuss}Discussion and conclusion}

The partial degradation of the Endothelial Surface Layer with neuraminidase yielding no discernible change in actin organization may seem puzzling in the first place, as the glycocalyx is directly involved in flow sensing of endothelial cells. However, the glycocalyx is a multicomponent layer, comprising different membrane-bound glycoproteins, proteoglycans and glycosaminoglycans having various functions \cite{Pahakis:2007cc}.  Neuraminidase specifically targets Sialic Acids (SA), which contribute to a large part of the overall thickness of the ESL and to its permeability \cite{Betteridge:2017ev}. However, SA are associated to glycoproteins that are not directly connected to the actin skeleton \cite{Zeng:2018ei}. In addition, neuraminidase is expected to leave intact glycosaminoglycans such as Heparan Sulfate \cite{Pahakis:2007cc}, a component of the ESL connected to the actin cytoskeleton through trans-membrane syndecans \cite{Zeng:2018ei}. Our observation therefore indicates that SA are not involved in actin remodelling under shear stress, and suggests that other members of the ESL connected to the  actin cytoskeleton, such as Heparan Sulfate, are not affected by the steric effect of neighboring SA in playing their role in glycocalyx-mediated mechanotransduction \cite{Florian:2003eb,Thi:2004df,Tarbell:2006er,Ebong:2014kc,Yen:2015db}.

Furthermore, our experiments indicate that HUVEC cultured under a steady shear flow corresponding to a wall stress of 0.4 Pa display long and flow-oriented F-actin fibers. This observation is consistent with previous works and shows that the well-established picture of the atheroprotective phenotype holds in a situation where the endothelial cells are strongly confined and organized in 3D into capillary-sized microchannels. 

It has been shown in previous studies that such an atheroprotective phenotype is lost in regions of disturbed or low shear flow, favoring {\it e.g.} lipoprotein uptake, adhesion of leukocytes \cite{Traub:1998dn} and  inflammation \cite{Baratchi:2017cc}. In order to have a complete picture of the mechanical factors controlling the atherosusceptibility of the endothelium, it is not only important to study EC under well-controlled flow conditions, but also to investigate how EC dynamically react to changes in such conditions, such as discussed in recent works related to how EC sense changes in flow direction \cite{Wang:2012ih,Wang:2013hb}.

Along this line of studying dynamical changes, our study demonstrates that, starting from a shear-adapted phenotype, HUVEC respond within one hour to a 5-fold lowering of wall shear stress down to 0.08 Pa by a marked rearrangement of the actin cytoskeleton, with stress fibers being disrupted/disorganized and actin being redistributed into the peripheral band. This is consistent with and complements previous observations of the actin organization of HUVEC under high and low stress stimuli obtained from vertical-step flow experiments \cite{Chiu:1998ib}. Our observation of a thickening of the nuclei, concomitant with actin remodeling under low shear, suggests a slight inflation of the cellular body. This is consistent with a decrease in the number of stress fibers, which we expect to be associated with a relaxation of the cellular internal stresses.  

Finally, we observe a trend for cells adapted under high shear to detach from the walls after long exposure to low shear, whereas HUVEC cultured at low shear from the beginning of the experiment are observed to be well adhered to the channel walls after 96h of culture. This suggests that the observed detachment trend is likely to be transient only, and might be associated with a reorganization of focal adhesions accompanying actin remodeling. This observation clearly calls for further investigation monitoring adhesion proteins \cite{Girard:1995hn} during flow reduction. However, such an observation, along with the above observations of actin rearrangement, strongly suggests that upon a five-fold decrease of the wall shear stress, HUVEC switch within a few hours from an atheroprotective to an atheroprone and more motile  phenotype \cite{Dai:2004kq}. 

While previous studies have highlighted differences in phenotypes in response to flow start-up \cite{Zeng:2014ji} or to various flow patterns applied from the start of the culture \cite{Chiu:2011iw,Zheng:2017kn,Dai:2004kq}, the present study reports on cytoskeleton reorganization of shear-adapted cells in response to a flow decrease. We note that such a response involves a reorganization of the actin skeleton that shares features (disruption of fibers, actin buildup at periphery of the cells) with what has been observed as a result of exposing endothelial cells to simulated microgravity \cite{Versari:2007bg}. This suggests that, in the context of vascular adaptation to space flight conditions, both microgravity and altered blood flow act in concert to induce phenotypic changes of endothelial cells.

\begin{acknowledgments}
We acknowledge the department of Material Sciences of the french Centre National d'Etudes Spatiales (CNES) for financial support. M.I is a PhD fellow of the France-Germany ``Living fluids'' Doctoral school.
\end{acknowledgments}

\bibliography{biblio}

\end{document}